\documentclass[traditabstract]{aa} 

\usepackage[dvips]{graphicx}          
\usepackage[comma,authoryear]{natbib} 
\bibpunct{(}{)}{,}{a}{}{,}            

\begin{document}

   \title{Convectively stabilised background solar models for local helioseismology}


   \author{H. Schunker
               \and
           R. Cameron
           \and
           L. Gizon
          }

   \institute{ Max-Planck-Institut f\"ur Sonnensystemforschung,
                  Max-Planck Str. 2, 37191 Katlenburg-Lindau,
                  Germany\\
                  \email{schunker@mps.mpg.de}
                               }

  \abstract
   {In local helioseismology numerical simulations of wave propagation are useful to model the interaction of solar waves with perturbations to a background solar model.  However, the solution to the equations of motions include convective modes that can swamp the waves we are interested in. For this reason, we choose to first stabilise the background solar model against convection by altering the vertical pressure gradient. Here we compare the eigenmodes of our convectively stabilised model with a standard  solar model (Model~S) and find a good agreement.}


   \maketitle

\section{Rationale}

One of the goals of local helioseismology is to interpret observations of the solar oscillations in the vicinity of sunspots and active regions. Kilogauss magnetic fields near the surface of the Sun have a strong effect on solar oscillations, which cannot be computed easily using perturbation theory. Numerical simulations of wave propagation may be the only realistic way to model these effects accurately. Two approaches have been proposed. The first one is to solve the equations of radiative MHD \citep[e.g.][]{Rempel09} in order to model the magnetic structures and the (naturally excited) seismic waves simultaneously. The second approach is to compute the propagation of waves through a prescribed magnetised background model \citep[e.g.][]{Cally1997,Cameron:2007p450,Cameron:2008p405}. While the first approach is more realistic, the second approach is less computer intensive.

Here, as a first step, we consider the problem of simulating the time evolution  of small-amplitude waves through a quiet Sun background atmosphere using the Semi-spectral Linear MHD code \citep{Cameron:2007p450}. In principle, any realistic model of the upper layers of the Sun, such as Model S \citep{JCD}, is convectively unstable.  It is however not possible to use Model S as a background model for an initial-value linear-wave calculation, since the convective modes will grow exponentially and quickly dominate the solution. These convective modes must be dealt with.

\section{The model}

We impose two main conditions for constructing a quiet-Sun model, which can be used in our simulations: (C1) it must be convectively stable and (C2) it must 
support oscillations which have solar-like eigenfrequencies and eigenfunctions.We choose to begin from Model S and to modify it so that the resulting model best satisfies these conditions.

There are various ways to achieve convective stability. For example, one may adjust the adiabatic exponent of the model. In this study, we choose to increase slightly  the pressure gradient such that $\partial_z p > c^2 \partial_z \rho$, where the sound speed, $c$, and the density gradient, $\partial_z \rho$, are retained from Model S. This guarantees that condition (C1) is met. The drawback is that  the p-mode kinetic energy densities are too small near the surface
compared to those of Model S and that the frequencies of the p modes are slightly too high.  We can compensate for this by changing the sound speed by a few per cent near the surface (details and justification discussed by \citet{Schunker09}. We arrive at a particular Convectively Stabilised Model (CSM), which
has eigenfunctions that are  as close as possible to Model~S while the eigenfrequencies remain within about 2\% of Model~S values.

\begin{figure*}[htbp]
\hspace{2cm}
\includegraphics[width=13cm,height=18cm]{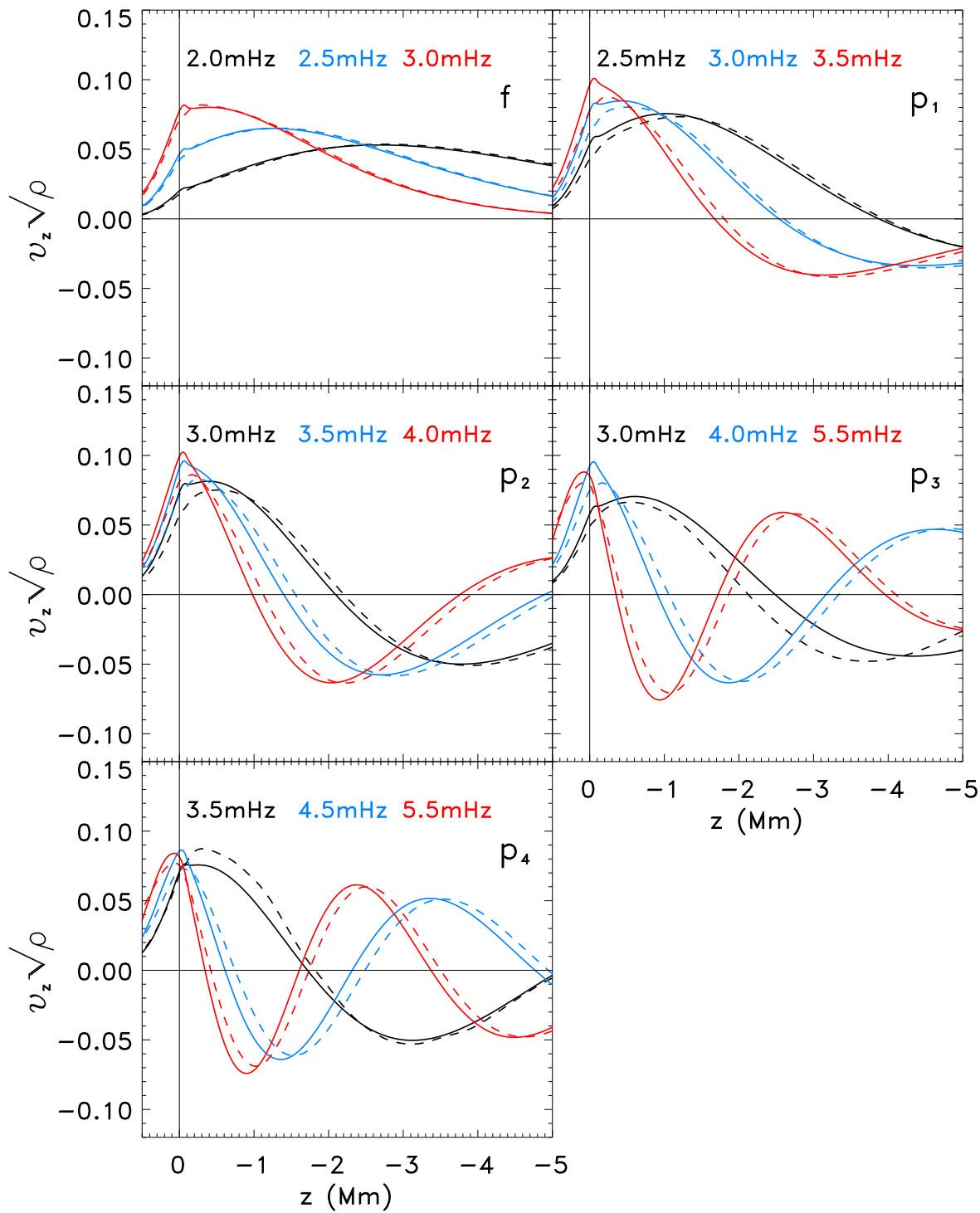}
\vspace{2cm}
\caption{
Plots of $v_z \sqrt{\rho}$ as functions of height, where $v_z$ is a vertical velocity eigenfunction and $\rho$ is density. The dashed curve is the convectively stabilised model (CSM) and the solid curve is Model~S.  Each panel shows eigenfunctions for one particular radial order as labelled at the indicated frequencies (black, blue, red).}
\label{smsa_efunc}
\end{figure*}
\begin{figure}[htbp]
\begin{center}
\includegraphics[width=8cm]{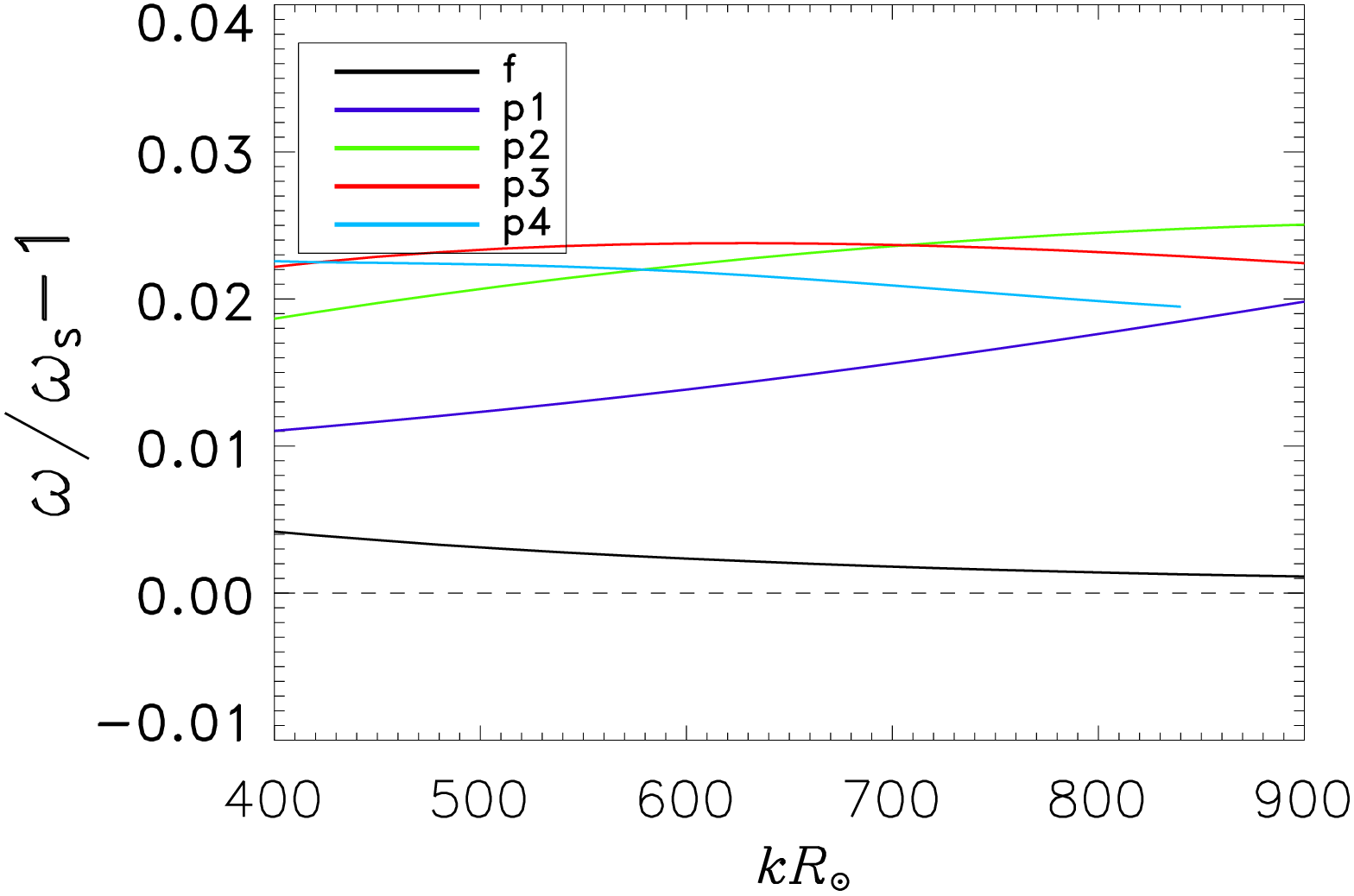}
\vspace{0.5cm}
\caption{The relative difference between  CSM eigenfrequencies ($\omega$)  and  Model~S eigenfrequencies ($\omega_S$) as a function of 
wavenumber, for modes $f$ through to $p4$.}
\label{efreqa}
\end{center}
\end{figure}

The vertical velocity eigenfunctions of  CSM were calculated numerically for each individual mode as described by \citet{Schunker09}. The 2D simulation box is 145~Mm in width and covers heights from $z=-20$~Mm to $z=0.7$~Mm (where $z=0$ is the base of the photosphere).  The model imposes damping layers above and below to minimise reflection from the boundaries. Due to the height of the box, we are limited to modelling modes with radial orders up to $4$.   Figure~\ref{smsa_efunc} shows the eigenfunctions for both CSM and Model S for each ridge, f through to p$_4$, at various frequencies.  Figure~\ref{efreqa} shows the relative difference between the CSM and Model S eigenfrequencies.

\begin{figure}[htbp]
\begin{center}
\includegraphics[width=8cm]{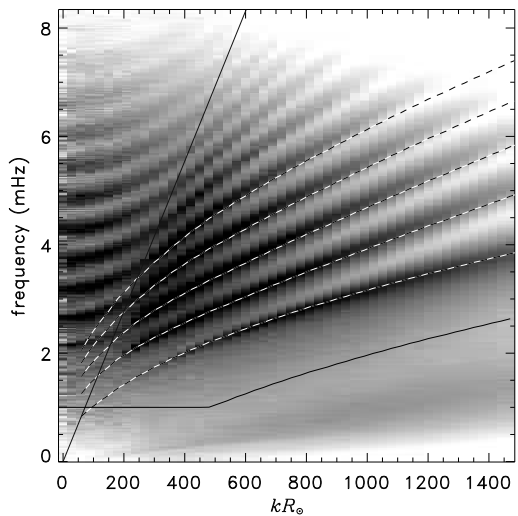}
\vspace{0.5cm}
\caption{The azimuthally averaged power spectrum of 8~hours of simulation data. The overplotted dashed curves are the eigenfrequencies of Model~S.  The solid line is the phase speed at the bottom of the box, above which we cannot expect to reproduce the solar spectrum. The lower solid curve excludes the gravity modes below, that were introduced by stabilising the background.}
\label{powspec}
\end{center}
\end{figure}

In addition to studying the eigenmodes of the model,
we wish to check that the power spectrum of oscillations compares favourably against that of the Sun. We extend the simulation to three dimensions ($145.77~\textrm{Mm} \times 145.77~\textrm{Mm} \times 27.5~\textrm{Mm}$) and introduce random sources 100~km below the surface. The sources are spatially uncorrelated and have an auto-correlation function $\exp(-t^2/2\tau^2)$ where $t$ is the correlation time lag and $\tau=5.5$~minutes. An additional attenuation term taken from \citet{GB02} is added in the momentum equation to model the finite lifetimes of the modes. The power spectrum of 8~hours of simulated data is shown in Figure~\ref{powspec}. The dashed curves are the eigenfrequencies of Model~S. Waves with phase speeds higher than $\omega / k = 5.86 \times 10^{6}$~cm~s$^{-1}$ (solid line) cannot expect to be modelled correctly because they hit the bottom of the simulation box.  When comparing with real solar observations, we find the agreement encouraging: for example, the mode asymmetries and location of the mode ridges compare well, but the mode amplitudes and line-widths could still be improved.

In short, we have constructed a background solar model which is convectively stable, solar-like, and can be used in local helioseismology studies. Opportunities to further improve this model remain and \citet{Schunker09} present a second convectively stabilised model with improved eigenfrequencies.

\begin{acknowledgements}
HS is supported by   the European
Helio- and Asteroseismology Network (HELAS), a major international collaboration
funded by the European Commission's Sixth Framework Programme.
\end{acknowledgements}

\bibliographystyle{aa}
\bibliography{schunker_HELAS_NA3}

\begin{thebibliography}{7}
\expandafter\ifx\csname natexlab\endcsname\relax\def\natexlab#1{#1}\fi

\bibitem[{{Cally} \& {Bogdan}(1997)}]{Cally1997}
{Cally}, P.~S. \& {Bogdan}, T.~J. 1997, \apjl, 486, L67+

\bibitem[{Cameron {et~al.}(2007)Cameron, Gizon, \&
  Daiffallah}]{Cameron:2007p450}
Cameron, R., Gizon, L., \& Daiffallah, K. 2007, Astronomische Nachrichten, 328,
  313

\bibitem[{Cameron {et~al.}(2008)Cameron, Gizon, \& Duvall}]{Cameron:2008p405}
Cameron, R., Gizon, L., \& Duvall, T.~L. 2008, Solar Physics, 51

\bibitem[{{Christensen-Dalsgaard} {et~al.}(1996){Christensen-Dalsgaard},
  {Dappen}, {Ajukov}, {Anderson}, {Antia}, {Basu}, {Baturin}, {Berthomieu},
  {Chaboyer}, {Chitre}, {Cox}, {Demarque}, {Donatowicz}, {Dziembowski},
  {Gabriel}, {Gough}, {Guenther}, {Guzik}, {Harvey}, {Hill}, {Houdek},
  {Iglesias}, {Kosovichev}, {Leibacher}, {Morel}, {Proffitt}, {Provost},
  {Reiter}, {Rhodes}, {Rogers}, {Roxburgh}, {Thompson}, \& {Ulrich}}]{JCD}
{Christensen-Dalsgaard}, J., {Dappen}, W., {Ajukov}, S.~V., {et~al.} 1996,
  Science, 272, 1286

\bibitem[{{Gizon} \& {Birch}(2002)}]{GB02}
{Gizon}, L. \& {Birch}, A.~C. 2002, \apj, 571, 966

\bibitem[{{Rempel} {et~al.}(2009){Rempel}, {Sch{\"u}ssler}, {Cameron}, \&
  {Kn{\"o}lker}}]{Rempel09}
{Rempel}, M., {Sch{\"u}ssler}, M., {Cameron}, R.~H., \& {Kn{\"o}lker}, M. 2009,
  Science, 325, 171

\bibitem[{{Schunker} {et~al.}(in preparation){Schunker}, {Cameron}, {Moradi},
  \& {Gizon}}]{Schunker09}
{Schunker}, H., {Cameron}, R., {Moradi}, H., \& {Gizon}, L. in preparation,
  \solphys

\end{thebibliography}
\end{document}